%% file: main.tex
\documentclass[aps,prl,twocolumn,showpacs,superscriptaddress,nobibnotes]{revtex4-1}

\input{header.tex}

\begin{document} 

\preprint{APS/123-QED}

\title{\LARGE\bf Rydberg-positronium velocity and self-ionization studies in 1T magnetic field and cryogenic environment}

\input{aegis_authors_list.tex}
\date{\today}
      
\begin{abstract}
We characterized the pulsed Rydberg-positronium production inside the \aegis{} (Antimatter Experiment: Gravity, Interferometry, Spectroscopy) apparatus in view of antihydrogen formation by means of a charge exchange reaction between cold antiprotons and slow Rydberg-positronium atoms. Velocity measurements on positronium along two axes in a cryogenic environment ($\approx$\SI{10}{\kelvin}) and in \SI{1}{\tesla} magnetic field were performed. The velocimetry was done by MCP-imaging of photoionized positronium previously excited to the $n = 3$ state. One direction of velocity was measured via Doppler-scan of this $n=3$-line, another direction perpendicular to the former by delaying the exciting laser pulses in a time-of-flight measurement. Self-ionization in the magnetic field due to motional Stark effect was also quantified by using the same MCP-imaging technique for Rydberg positronium with an effective principal quantum number $n_{eff}$ ranging between 14 and 22. We conclude with a discussion about the optimization of our experimental parameters for creating Rydberg-positronium in preparation for an efficient pulsed production of antihydrogen.
\end{abstract}

\pacs{32.80.Rm, 36.10.Dr, 78.70.Bj}
\maketitle{}

\section{Introduction}

Antimatter has been thoroughly studied for almost a century now, first theoretically, when the antiparticle of the electron - the positron ($e^+$) -  emerged from Dirac's equations \cite{Dirac28}, and a few years later also experimentally, when Anderson observed it for the first time \cite{Anderson32}. Since then, positronium (Ps), the bound state of a positron and an electron, was predicted and discovered \cite{Moho34,deutsch_ps:51}. Increasingly precise values for higher excited states of Ps plus their annihilation and de-excitation rates were calculated using methods from quantum-electrodynamics (QED) \cite{Alekseev58,Alekseev59,Caswell79}, and eventually experiments showed many of these features to be correct in the limits of reachable precision \cite{cassidy_paschen:11,cassidy_hyp:12,aegis_meta:18,chu_mills:82,cassidy_review:18}. 


Of special interest is antihydrogen (\Hbar{}), the first observed anti-atom in a laboratory \cite{LEAR:Hbar1995}. The first \textit{cold} antihydrogen was produced by direct mixing of antiprotons and positrons \cite{athena_nat:02,atrap_hbar:02}. 
The \aegis{} experiment (Antimatter Experiment: Gravity, Interferometry, Spectroscopy) \cite{aegis_spsc:18} has, between others, the objective to produce antihydrogen to experimentally probe the antimatter gravitational acceleration. In particular, the current goal of \aegis{} is to demonstrate \textit{pulsed} cold antihydrogen production via charge-exchange reaction \cite{charlton:90}:
\begin{equation} \label{eq:charge}
    Ps^* + \bar{\hbox{p}}\rightarrow e^- + \bar{\hbox{H}}^*,
\end{equation}
where Ps$^*$ is a positronium atom excited to a Rydberg state, hence a state with high principal quantum number $n$, the symbol $\bar{\hbox{p}}$ denotes an antiproton, $e^-$ the standard electron and $\bar{\hbox{H}}^*$ is a Rydberg-antihydrogen atom. The greatest advantage using the charge-exchange reaction is the scaling of its cross-section with $n^4$. However, as the experiment has to be executed in a strong magnetic field of \SI{1}{\tesla}, the addressed Rydberg state must not be too high in order to avoid Ps self-ionization. Furthermore, the Ps has to be in the right velocity regime because fast Ps components have a small cross-section \cite{krasnicky_pra:16} and their contribution to the charge exchange reaction is negligible. Here, we neglect antiproton velocities, since these are generally two orders of magnitude smaller than for Ps, which allows in our case to monitor only the Ps-velocity distribution in order to characterize the charge exchange reaction. From the study of the charge-exchange cross-section performed in Ref. \cite{krasnicky_pra:16} it follows that the useful Ps-velocities range from zero up to $\sim\SI{1.3e5}{\meter\per\second}$, where the upper limit is defined by the heavily decreasing cross-section for the addressed Rydberg-state at higher velocities.

In this work we have produced ortho-positronium via implantation of positron bunches into a positron/Ps converter kept at \SI{10}{\kelvin} in the \SI{1}{\tesla} field of the \aegis{} apparatus. Positronium emitted into vacuum has been laser-excited via the two-step transition $1^3$S $\rightarrow3^3$P $\rightarrow$ Rydberg levels \cite{castelli_psexc:08, aegis_neq3:16}. The velocity of produced Ps and the fraction of Rydberg-Ps surviving self-ionization caused by motional Stark effects in the magnetic field \cite{castelli_psexc:08} have been characterized in view of the optimization of the charge-exchange cross-section. The characterization has been performed via MCP-imaging \cite{aegis_MCP:19} of ionized $3^3$P-Ps and Rydberg-Ps, respectively. In this paper we present the experimental results of the \Ps{}-source characterization and the techniques by which we adjusted the system's parameters for creating Rydberg-Ps in view of antihydrogen production.

\section{EXPERIMENT}

\begin{figure}[thpb]
      \centering
      \includegraphics[width=1\linewidth]{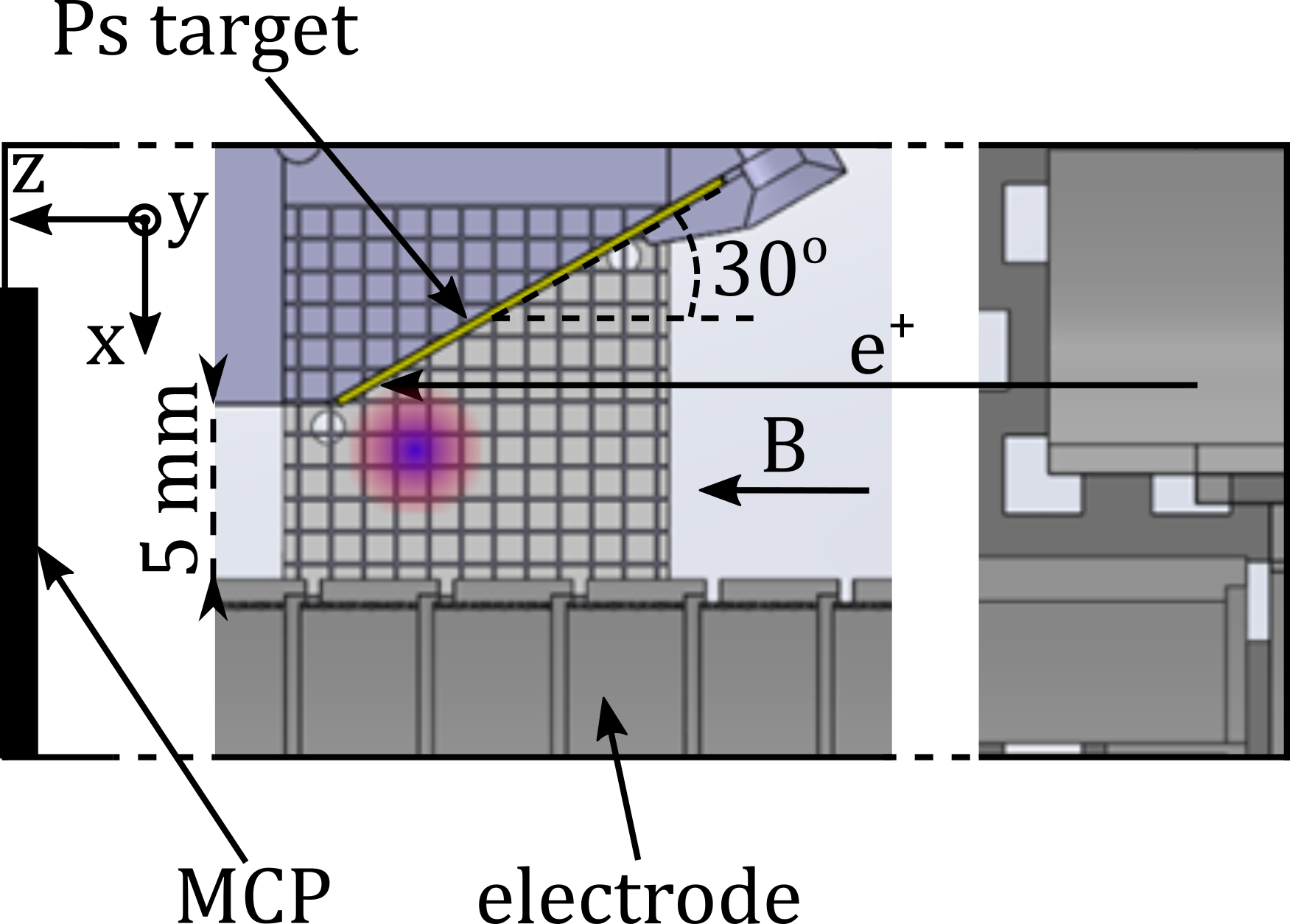}
      \caption{Sketch of the \Ps{}-excitation area. One can see the \macor{} screen in the back, covered with a mesh-grid, the positron/Ps converter target and the trap electrodes. Note the \SI{5}{\milli\meter} gap between the bottom of the converter target and the top of the electrodes which enables the \Ps{} velocimetry. The two-colored spot marks the laser position of an UV-beam (purple), superimposed with an IR-laser (red) used for Ps-excitation as specified in the text. The lasers are centered at \SI{-3.2}{\milli\meter} in z-direction and \SI{1.7}{\milli\meter} in x-direction from the lower edge of the target.}
	\label{fig:aegis_scheme}
   \end{figure}

The \aegis{} apparatus, the environment for the presented measurements, consists of: a positron system, able to accumulate up to some \SI{e7}{} positrons from a 25\,mCi $^{22}$Na source within few minutes and to magnetically transport them to a kicker, an isolated segment of the beam tube where a high potential pulse can be applied to (like this, the positrons can reach a few \SI{}{\kilo\electronvolt} kinetic energy); a positron/positronium converter placed in the \SI{1}{\tesla} cryogenic environment ($\approx$\SI{10}{\kelvin}) where antihydrogen will be formed; a laser system to excite positronium to Rydberg states; an antiproton trapping/cooling system (not operated for the measurements presented here). The used converter was a nanochanneled silicon target similar to the one reported in Ref. \cite{mariazzi_prl:10} with a channel diameter of about \SI{10}{\nano\meter} and a depth of $\sim$\SI{1.5}{\micro\meter}. It was tilted by \SI{30}{\degree} against the positron beam axis, while the implantation energy of positrons was set via the kicker to \SI{4.6}{\kilo\electronvolt} in order to allow produced Ps atoms to cool enough before they are emitted into vacuum. The whole assembly for Rydberg-Ps production and excitation is sketched in Fig. \ref{fig:aegis_scheme}. For the measurements presented here, a few \SI{e6}{} positrons per bunch have been used. Positronium emission was observed to be slightly less than \SI{10}{\percent} with respect to the implanted positrons \cite{Caravita2019}. The Ps yield was mainly limited by the adsorption of contaminants from vacuum over time \cite{cooper16}, which could not be removed from the converter due to a failure in the dedicated target heater, whose repair would have caused a too long interruption for parallel antihydrogen measurements in the \SI{1}{\tesla} region.

The positronium excitation is achieved by two synchronized and superimposed laser pulses. The first one enabling the $1^3$S$\rightarrow3^3$P transition is a \SI{1.5}{\nano\second} long UV-pulse, tunable from \SIrange{204.9}{205.2}{\nano\meter} at \SI{43}{\micro\joule} with a bandwidth of \SI{118}{\giga\hertz}. The second is for the transition to Rydberg states, namely an IR-pulse, \SI{3}{\nano\second} long and tunable from \SIrange{1671}{1715}{\nano\meter} at \SI{1.6}{\milli\joule} with a bandwidth of \SI{430}{\giga\hertz}. 
In addition, a \SI{1064}{\nano\meter} IR-pulse at \SI{30}{\milli\joule} can be delivered in order to selectively photoionize all Ps in the $3^3$P state. The positron \cite{aegis_nimb:15} and laser \cite{aegis_neq3:16} systems are described in detail elsewhere. 

The electrodes of the antihydrogen production trap, placed below the Ps-excitation area and partially visible in Fig.\ref{fig:aegis_scheme}, allow to store antiprotons during the procedure. The maximum antihydrogen production within the magnetic field $\vec{B}$ (parallel to the z-axis) will be achieved by steering the highest Rydberg-\Ps{} flux originating from the target towards that trap, and by optimizing the Ps-state as well as the laser-selected Ps velocity regime in the x-y plane. At the top of these trap electrodes is an opening covered by a meshed grid which allows Ps to enter. The exciting laser pulses are aligned parallel to the \pos{}/\Ps{} converter target, thus along the y-axis. In \fig{aegis_scheme} we indicate the position of the UV and IR laser-beams which was used throughout the whole presented measurement series. A \macor{}- screen has been placed in the proximity of the target in the x-z plane in order to image the laser pulses and monitor their position. For the measurements presented here, the central position of the laser was set to \SI{-3.2}{\milli\meter} in z-direction and \SI{1.7}{\milli\meter} in x-direction from the lower edge of the target. 

The laser timing was referenced to the prompt annihilation peak of implanted positrons, which we defined as $t=0$. For this, the positron annihilation radiation was detected with scintillator slabs (EJ200) coupled to a photomultiplier tube (PMT). Additionally, we acquired the electric signal on the MCP frontface which was generated immediately after irradiation with scattered UV light. Both the positron peak and the laser peak were absolutely calibrated to each other as described in Ref. \cite{aegis_SLOPOS15:2019}.
Using the MCP imaging technique, a single image could be acquired for one specific setting at the same repetition rate as the \aegis{} positron system was delivering positron bunches, namely 1 minute per bunch. If a setting was repeated up to five times, a full measurement series required approximately $5 \times 60$ different settings $\times\,\SI{1}{\minute} = \SI{300}{\minute}$.

The characterization of the Ps velocity profiles and the fraction of Rydberg-Ps surviving self-ionization was studied by performing three different types of experimental procedures: (i) timing scans, (ii) Doppler scans and (iii) Rydberg self-ionization scans. 
\begin{itemize}
    \item[(i)] The timing scans focused on the determination of the Ps-velocity component aligned with the vertical x-axis by means of photoionization of Ps atoms via a two-step optical transition and a subsequent measurement of the intensity distribution of released positrons. The timing scans were performed by changing the delay between positron implantation into the target and the trigger for laser excitation. Delays from \SI{-5}{\nano\second} up to \SI{265}{\nano\second} were applied to both the UV-pulse exciting \Ps{} via $1^3$S$\rightarrow3^3$P, operating at resonance ($\lambda=\SI{205.045}{\nano\meter}$, taken as a reference in the following), and to the IR-pulse at \SI{1064}{\nano\meter} in order to photoionize selectively the fraction of \Ps{} in the $n=3$ state. This fraction goes up to \SI{15}{\percent} of the whole Ps-cloud being emitted into vacuum \cite{aegis_neq3:16}. 
    The released photo-positrons were guided by the homogeneous \SI{1}{\tesla} magnetic field towards a Micro-Channel Plate coupled to a phosphor screen (MCP Hamamatsu F2223 + Phosphor screen P46) and imaged by this assembly \cite{aegis_MCP:19}. The front-face of the MCP was biased to \SI{-180}{\volt}, the gain voltage between the two stages of the MCP has been set to \SI{1120}{\volt}. The intensity distribution of the positrons released by Ps-photoionization vs. laser-pulse delays was thus acquired. Using the known x-position of the Ps-origin, the obtained distribution can be converted to a discrete velocity distribution along the x-axis, $v_x$.
    
    \item[(ii)] The Doppler scans \cite{cassidy_silica:10,aegis_neq3:16} are performed in order to determine the Ps-velocity distribution parallel to the target, i.e. $v_y$. These scans were performed for fixed laser-delays (once \SI{23}{\nano\second} and \SI{31}{\nano\second} for a second series), but with linearly adjustable UV-wavelength between \SI{204.900}{\nano\meter} and \SI{205.200}{\nano\meter} in steps of \SI{0.005}{\nano\meter}. Like this, one can investigate the $1^3$S$\rightarrow3^3$P transition in the vicinity of resonance. The IR laser was again set to \SI{1064}{\nano\meter} as for (i) in order to produce photo-positrons. 
    
    \item[(iii)] The self-ionization scans reveal the fraction of Rydberg-Ps surviving the magnetically induced motional Stark field as a function of the effective principal quantum number $n_{eff}$, a quantum number characterizing the Rydberg states in fields that will be defined later. These scans were done by changing the IR-wavelength that is inducing the transition from $3^3$P$\rightarrow$Rydberg. It was tuned from \SI{1671}{\nano\meter} to \SI{1715}{\nano\meter}, addressing effective Rydberg states between 14 and 22, while the UV-laser was kept unchanged and at the reference wavelength. Both laser pulses were synchronously delayed by \SI{25}{\nano\second}. The internal MCP gain voltage was increased to \SI{1200}{\volt} in order to be more sensitive to the emerging positrons from self-ionization effects. This enhancement of sensitivity is necessary, because a low-intensity signal is expected for Ps self-ionization at higher IR-wavelengths. The intensity of the image on the MCP was thus acquired as a function of the effective n-state of Ps.
\end{itemize}

\section{RESULTS AND DISCUSSION}

\begin{figure}[thpb]
      \centering
      \includegraphics[width=1 \linewidth]{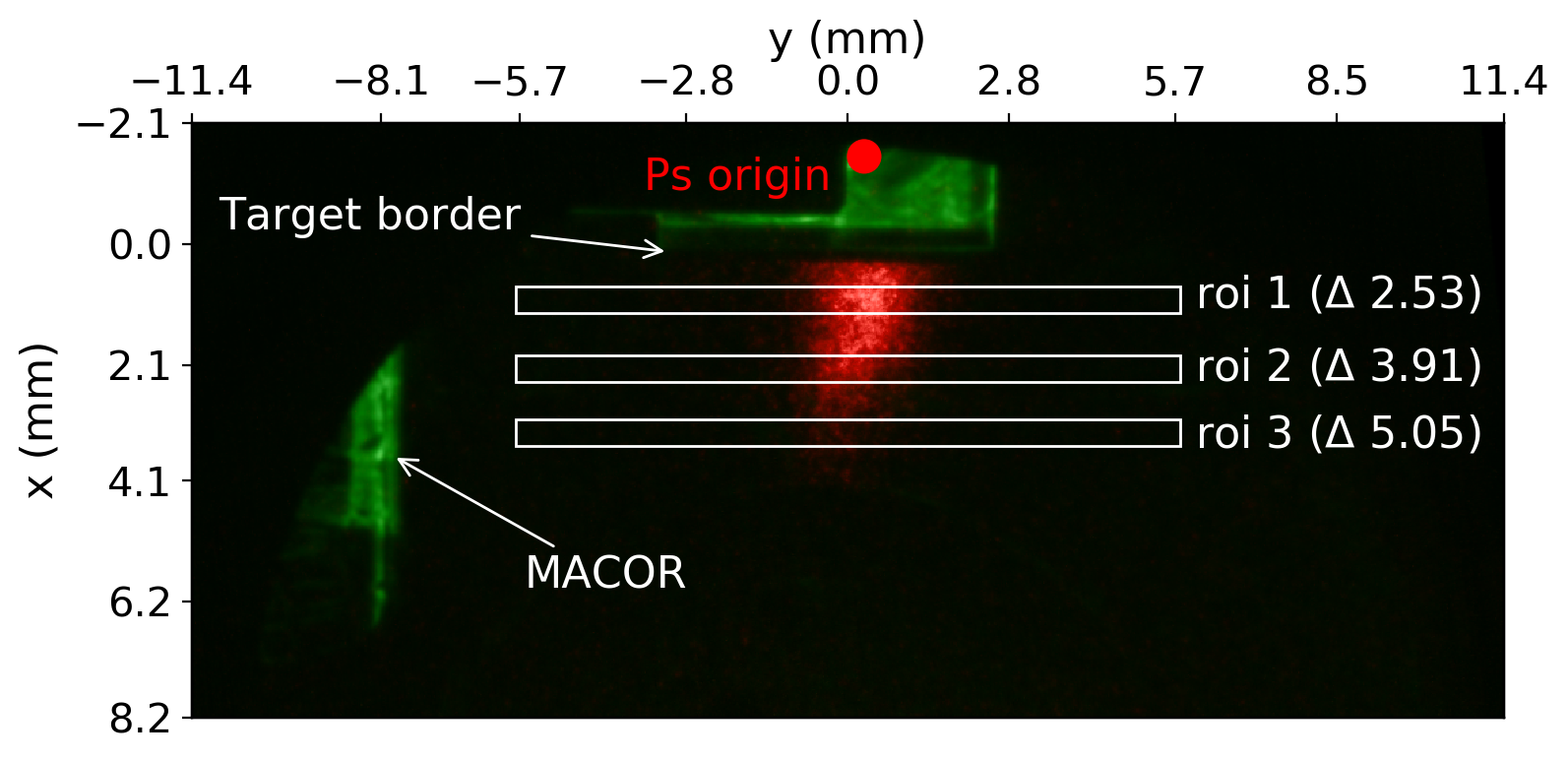}
      \caption{Combination of an electron reference image (green) and a \Ps{} photoionization image (red): Positronium is emerging from the target border and gets instantaneously ionized when the laser pulses are sent. Unbound positrons are then guided towards the MCP, producing a light-spot. We define regions of interest (\emph{roi 1, roi 2, roi 3}), which have a width of 20 pixel (\SI{0.46}{\milli\meter}) and a length of 500 pixel (\SI{11.5}{\milli\meter}). The window \emph{roi 1} is \SI{2.53}{\milli\meter} away from the Ps origin, \emph{roi 2} \SI{3.91}{\milli\meter} and \emph{roi 3} \SI{5.05}{\milli\meter}. We sum all the intensities within the windows and subtract a background obtained from the averaged signal of the last six delays where no Ps-photoionization could be observed.}
    \label{fig:roi}
    \end{figure}

The measurements produce a photo-positron induced image on the MCP+phosphor screen assembly, acquired by a CMOS camera (Hamamatsu C11440-22CU). An example for a signal after photoionization is shown in \fig{roi}. One can see the ionized positronium cloud (red), emerging from below the border of the target. Note that here for visual verification of the internal alignment the image was overlayed with an electron-image (green), where the MCP front-face was charged with +\SI{180}{\volt}, which attracted in turn photo-electrons released from all surfaces after irradiation with the UV-laser \cite{AEGIS:MCP1T_19}. As a result, one can see the target border plus its holder on the top and the \macor{}-screen on the left. The laser approaches from the right in this view. The back of the target seems also to emit photo-electrons, which we attribute to multiple scattering of the UV-laser. For analysis, only the positron-image was taken into account.

The transverse spatial profile and energy-distribution of both laser-pulses, which are well-described by a Gaussian, determine the absolute amount of ionized and thus detectable Ps-atoms. A fraction of the photo-released positrons, constrained to move along the magnetic field lines in z-direction, is hidden behind the target, the holder or just out of the MCP-range and is therefore not available for the analysis. The UV beam has a full-width-half-maximum (FWHM) of approximately \SI{4.4}{\milli\meter}, which at the same time defines the border of a sufficiently high fluence guaranteeing saturation of the Ps excitation across the illuminated surface \cite{aegis_MCP:19}. The FWHM of the IR-laser is about $\sim4-$\SI{5}{\milli\meter}. For analysis, we choose several rather narrow windows in the x-y-plane on the MCP-image, of which we will present three specific ones labeled \emph{roi} 1, \emph{roi} 2 and \emph{roi} 3 in Fig. \ref{fig:roi}. The windows have an extent of 20 x 500 pixel (0.46 x \SI{11.5}{\milli\meter}) and cover the visible region of \SI{5}{\milli\meter} along the x-axis.
For each \emph{roi}, the integrated signal of ionized Ps along the z-axis depends on the geometrical overlap of the window with the laser beam at the instant of excitation. This implies that one can effectively compare different delays for the same \emph{roi}, while other windows have different overall intensities.
Generally, a narrow window is preferable since the broadness of it has an influence on how precise a specific velocity can be estimated in the case of timing scans. Furthermore, in the region close to the target border edge-effects (e.g. inhomogeneous fields) might influence the analysis.

As a first step in the experimental procedure, we had to place correctly the position of the positron-beam incident on the converter, which we used as reference for the Ps-origin. For this, we have measured the dimension of the positron-spot in the center of the MCP, which was found to be \SI{0.46}{\milli\meter}. We then have moved up the positron implantation point by adjusting the current of a vertical correction coil in the  positron-beamline, which did not influence the shape of the positron spot. A linear correlation of the vertical x-coordinate and the current in the correction coil was observed. By means of this calibration we were able to steer the positron implantation point to a fixed, hidden position on the positron/Ps converter target which serves as zero. The distances between this point and the center of \emph{roi 1}, \emph{roi 2} and \emph{roi 3} were found to be $x_0=\SI{2.53+-0.38}{\milli\meter}$, \SI{3.91+-0.38}{\milli\meter} and \SI{5.05+-0.38}{\milli\meter}, respectively. The systematic errors on $x_0$ have been estimated by Gaussian error propagation, arising from the positron spot-size (\SI{+-0.23}{\milli\meter}), the calibration measurement (\SI{+-0.2}{\milli\meter}) and the broadness of the analysis window (\SI{+-0.23}{\milli\meter}). The highest Ps flux has been assumed to originate from the center of the positron implantation spot.

\subsection{(i) Timing scans}

\begin{figure}[thpb]
    \centering
    \includegraphics[width=1 \linewidth]{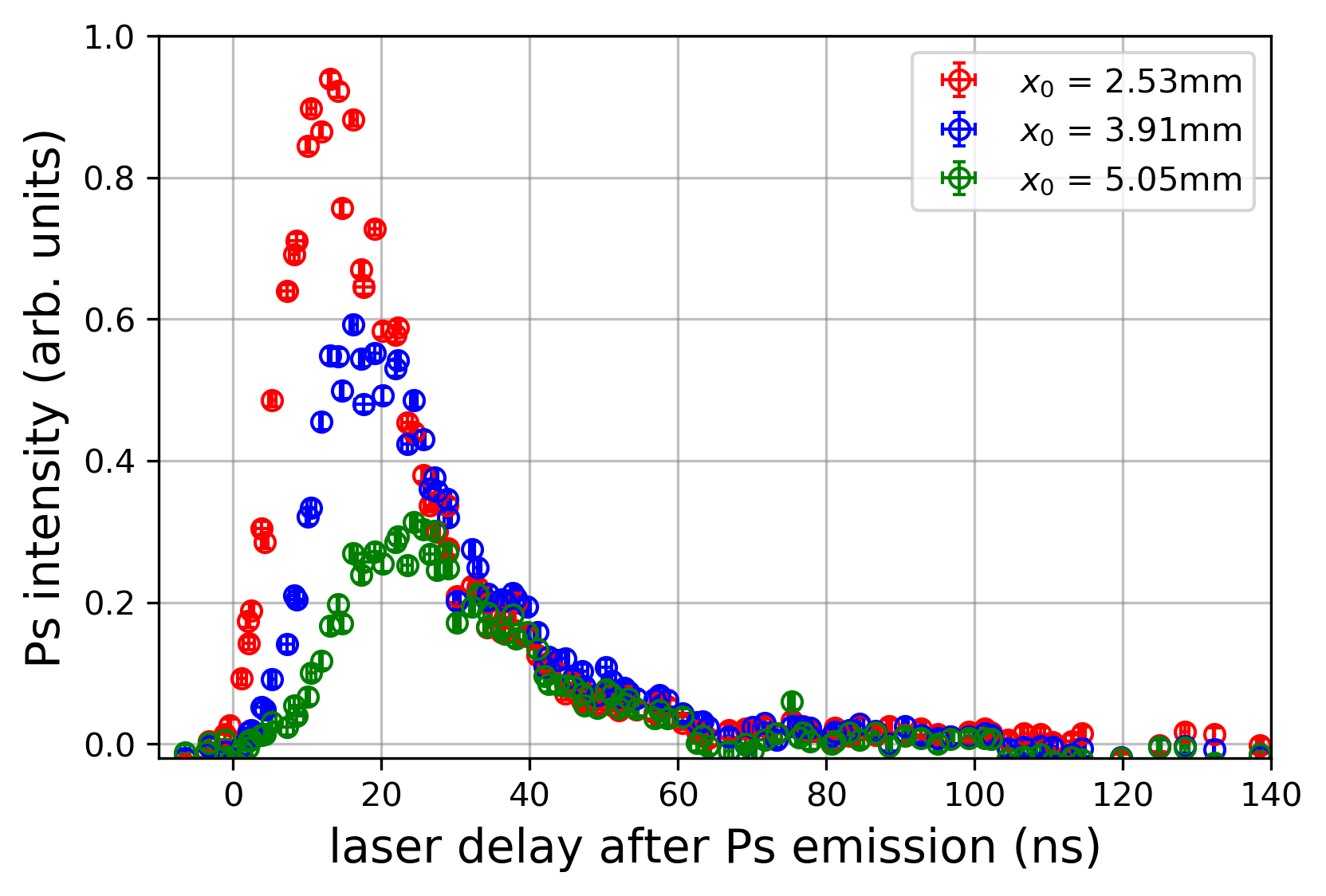}
    \caption{Observed positronium intensity vs. lasers delay for three different windows (\emph{roi} 1,2,3). The distributions shift towards later times for more distant windows and at the same time decrease in intensity due to the limited geometrical overlap with the laser.}
    \label{fig:timing_pos4}
\end{figure}

The timing scans were repeated three times, so that an average image could be calculated for each delay. A background obtained from the averaged signal at very long delays with no evident photoionization signal was subtracted from the raw data-points' amplitudes. We then obtain asymmetrical distributions displaying the intensities of observed Ps over different delays, see \fig{timing_pos4}. After \SI{140}{\nano\second}, there is no evident signal anymore. We find that the distributions shift towards later times for windows that are more distant from the Ps origin, which is what one would expect. Furthermore, the peak intensities decrease with increasing distance, which is the net effect of the laser-beam selecting a fraction of the expanding Ps cloud.

\begin{figure}[thpb]
    \centering
    \includegraphics[width=1\linewidth]{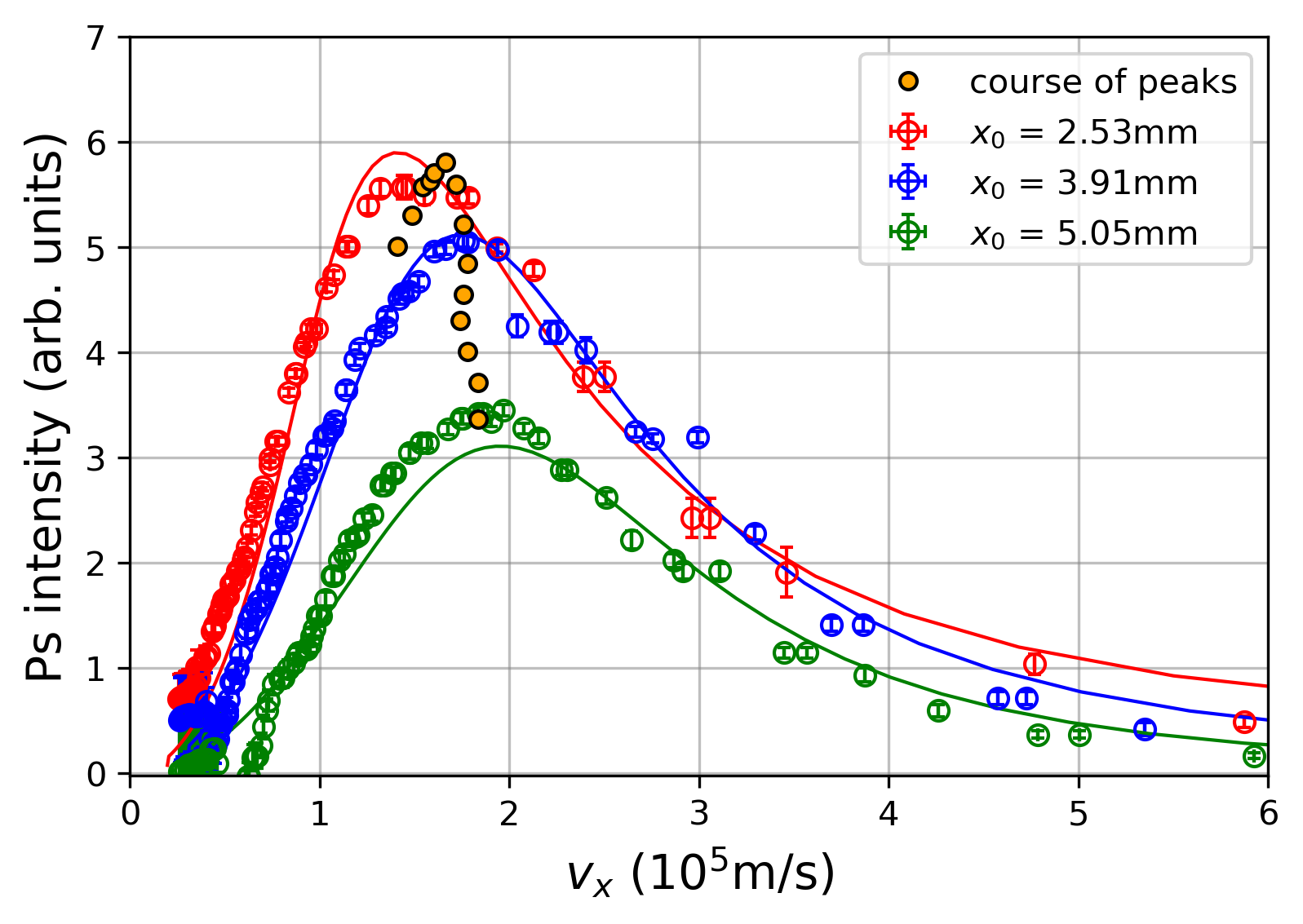}
    \caption{Distribution of Ps velocities along the x-axis as obtained from the timing scan of Fig. \ref{fig:timing_pos4} after the treatment specified in the text. The solid lines are obtained from a simulation which is specified in the appendix. The peak-velocities have been marked with orange spots not only for the defined analysis windows, but also for additional windows with the same extent, sliding in steps of 10 pixel from the target border towards the trap electrodes.} 
    \label{fig:velo_distribution}
\end{figure}

By using the relation $v_x = x_0/t$, we can translate the Ps time-of-flights $t$ into velocities using the distance $x_0$ between \Ps{} origin and the analysis windows introduced above. The velocity distributions were obtained by multiplying the observed Ps distributions of Fig. \ref{fig:timing_pos4} with the term $t\cdot e^{t/\SI{142}{\nano\second}}$, i.e. by using the Jacobian determinant for converting the observed space distribution into a velocity distribution and correcting for the natural decay of ground state ortho-\Ps{} as is usually done for \Ps{}-TOF measurements (\cite{PsTOF:87,Deller:TOF:2015}). Differently to these references, a single laser position was sufficient to resolve the velocity distribution in our highly magnetic environment, especially since a single MCP image provided many detection distances $x_0$ at the same time.
We then apply a sliding average in order to smooth the curves and increase the statistics especially for the less intense low velocity components (corresponding to long delays).
The result is reported in Fig. \ref{fig:velo_distribution}. The solid lines are the outcome of a simulation aiming to qualitatively model the performed experiment (more details are given in appendix). All distances investigated result in a velocity distribution peaking between \SI{1.4e5}{\meter\per\second} and \SI{1.8e5}{\meter\per\second} (orange circles), where more distant windows have higher velocities. 

We attribute this to a combination of two effects. The first is simply a geometrical effect, because for more distant \emph{rois} the contribution of Ps components aligned with the z-axis is decreasing. As a consequence, the cleanest $v_x$-distribution is obtained for the most distant window. 
The second effect is a permanence time of Ps in the nanochannels ($t_{perma}$), because in reality Ps emission from the nanochannels is not instantaneous \cite{cassidy_perman:10}. The real Ps time-of-flight rather is $t_{f}= t-t_{perma}$, where $t$ is the measured time elapsed from positron implantation to Ps ionization. The permanence time $t_{perma}$ depends both on the positron implantation depth and the nanochannel diameter \cite{mariazzi_prl:10}. 
For closer \emph{rois}, the presence of a permanence time yields a non-negligible overestimation of the generally smaller time-of-flights and, consequently, to an underestimation of Ps velocities. 
Increasing the \emph{rois}' distance from the Ps origin minimizes this disturbing effect of the permanence time, rendering together with the first effect the most distant window to be the most trustworthy. 
In Fig. \ref{fig:velo_distribution}, we report the most probable velocity as a function of the \emph{roi}-distance (course of peaks) showing that the peak velocity stabilizes for distances greater than $\approx\SI{4}{\milli\meter}$.

In the light of this discussion, we use the velocity-distribution obtained from the window at \SI{5.05}{\milli\meter} (green curve in Fig. \ref{fig:velo_distribution}) as reference. It has a most frequent velocity component $v_{x,max}$ at \SI{1.80+-0.25e5}{\meter\per\second}. The given error was derived from the width of the window that we used for the sliding average.

\subsection{(ii) Doppler scans}

\begin{figure}[htbp]
  \centering
    \includegraphics[width=\textwidth]{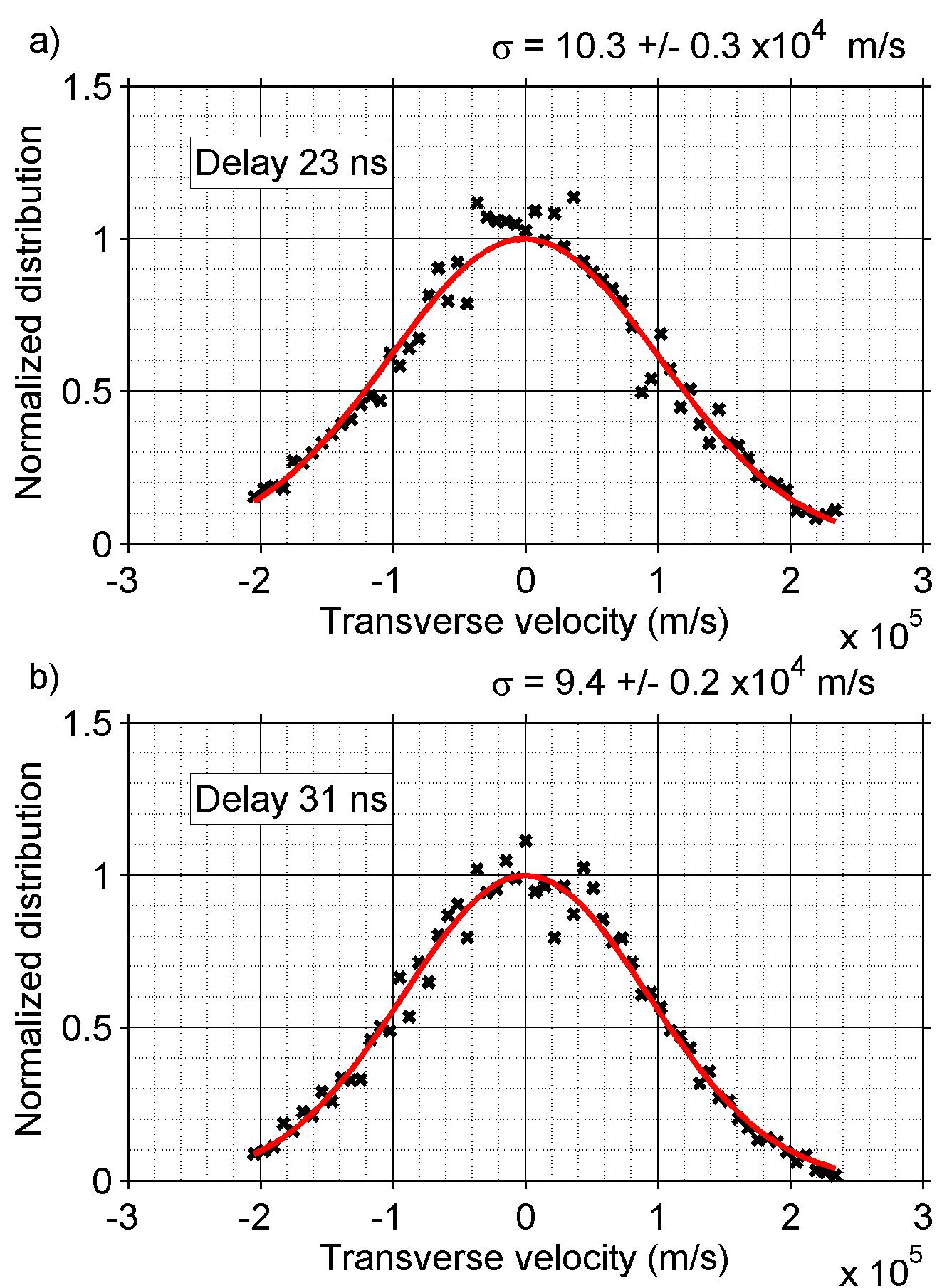}
  \caption{a) Doppler scan at \SI{23}{\nano\second} delay b) Doppler scan at \SI{31}{\nano\second} delay: Both distributions have a peak at $v_y=\SI{0}{\meter\per\second}$ corresponding to $\lambda = \SI{205.045}{\nano\meter}$.}
  \label{fig:velo_Doppler}
\end{figure}

The Doppler scans were repeated five times so that we could average the images per wavelength. The $v_y$-distribution has been extracted from the photoionization images in the following way. Within the reference window \emph{roi} 3, all intensity values of one averaged image have been summed up and a constant background, i.e. the average value of a region on the MCP image where no Ps can be photoionized, was subtracted. Repeating this for all wavelengths, one obtains the photoionization-signal dependent on the UV-laser setting. Furthermore, the wavelength can directly be expressed as the \Ps{} velocity-component $v_y$ propagating parallel/antiparallel to the laser beam by using the Doppler relation:
\begin{equation} \label{eq:Doppler}
	v_y = c\bigg(1 - \dfrac{\lambda}{\lambda_{r}}\bigg)
\end{equation}
where $c$ is the speed of light in vacuum and $\lambda_{r}=\SI{205.045}{\nano\meter}$ the reference wavelength for the excitation of the $n=3$ manifold in our experimental conditions.
In Fig. \ref{fig:velo_Doppler} the summed intensities are shown as a function of the velocity $v_y$ as well as Gaussian fits, from which the sigmas of the distributions have been extracted. 
Doppler scans strictly distinguish between velocities propagating towards the light source and those moving away from it. This is expressed by  positive and negative velocities in Fig. \ref{fig:velo_Doppler}, respectively. Note that this measurement of the $v_y$-distribution is independent from the result of timing scans, which led to the $v_x$-distribution. 
The plot labeled a), with a laser delay of \SI{23}{\nano\second}, has a sigma of \SI{1.03+-0.03e5}{\meter\per\second}. The plot labeled b), with a set delay of \SI{31}{\nano\second}, has a sigma of \SI{0.94+-0.02e5}{\meter\per\second}. These velocities correspond to the mean velocity of the distribution aligned with the y-axis. For the greater delay, the velocity-spread of positronium being in the visible region on the MCP seems a bit decreased, which could be due to faster Ps components having moved out of view or due to non-isotropic Ps-emission from the target. 

The best UV-wavelength for \aegis{}' setup is the one where the greatest fraction of Rydberg-Ps is moving towards the meshed grid at the top of the \Hbar{}-production trap. With the current alignment of laser, target and trap this is just the $v_y=0$ component, i.e. $\lambda = \lambda_{r}$.

\subsection{(iii) Self-ionization scan of Rydberg-Ps}

Rydberg-\Ps{} has a strong dipole moment scaling with the principal quantum number squared. Hence, it is sensibly affected by electric fields despite the general neutrality of Ps. This, on one hand, enables motional control for example via Stark deceleration as described in Ref. \cite{hogan_rydrev:16}, but on the other hand it also puts constraints to its production in strong magnetic fields. In such, the motion of Rydberg-\Ps{} induces an electrical field following $\vec{F}_{mot} = \vec{v}\times\vec{B}$. This is the so called motional Stark effect and its strength depends on the velocity components perpendicular to the magnetic field. Here, this is only $v_x$, because with the laser set to resonance we selected $v_y\approx\SI{0}{\meter\per\second}$. The presence of this electric field can cause self-ionization of the moving Ps-atom. In particular, the minimal electric field causing ionization on some Ps-states, usually called the \textit{ionization threshold}, generally depends on a principal quantum number $n$ \cite{castelli_psexc:08,Gallag05} and  can be written in our case as:
\begin{equation} \label{eq:Flimitraw}
    F_{limit}(n) = \dfrac{E_{Ps}}{9ea_{0}}\cdot\dfrac{1}{n_{eff}^4}
\end{equation}
Here, $E_{Ps}$ is the Rydberg energy for positronium (equal to $\SI{6.8}{\electronvolt}$), $e$ is the electric charge and $a_{0}$ is the standard Bohr-radius, while $n_{eff}$ represents an effective quantum number depending only on the wavelength $\lambda$ characterizing the $n=3\rightarrow\:$Rydberg transition through the well-established Rydberg-formula for hydrogen-like systems:
\begin{equation} \label{eq:Rydbergformula}
    \dfrac{hc}{\lambda} = E_{Ps}\bigg(\dfrac{1}{(n=3)^{2}} - \dfrac{1}{n_{eff}^{2}}\bigg)
\end{equation}

This reformulation is possible since the distribution of Ps-Rydberg-states resembles a continuum rather than well-separated single states as the classical formula would imply. In fact, the presence of the motional Stark effect is highly relevant for Rydberg-excited Ps-atoms: it destroys the axial symmetry and determines via mixing of $\ell$ and $m$ sub-states the spread of energy levels for the $n$-manifold. It also determines the interleaving of nearby $n$-manifolds, finally leading to a quasi–continuum structure of energy levels \cite{castelli_psexc:08,castelli:2012}. This also has the consequence that the efficiency of the transition $n=3\rightarrow\:$Rydberg is essentially dominated by the IR-laser’s bandwidth of \SI{430}{\giga\hertz}.

By using the relation in Eq. \ref{eq:Rydbergformula}, one finally obtains the self-ionization limit as a function of the IR-excitation wavelength $\lambda$:
\begin{equation} \label{eq:Flimit}
	F_{limit}(\lambda) = \dfrac{E_{Ps}}{9ea_{0}}\cdot\bigg(\dfrac{1}{9}-\dfrac{hc}{E_{Ps} \lambda}\bigg)^2,
\end{equation}
and from the relation $v_{x,limit} = F_{limit}/B$, with $B=\SI{1}{\tesla}$, the corresponding limiting velocity that is used in the following analysis of experimental data.\\

Also in the case of Rydberg-Ps self-ionization produced positrons can be imaged on the MCP, although the number of detectable particles turns out generally smaller. Therefore, unlike before, we used a larger analysis window ranging from the target border to \emph{roi} 3 for a better signal intensity, and normalized it with the corresponding signal from the measurement at the lowest possible wavelength ($\lambda=\SI{1671}{\nano\meter}$). At the same time, one can calculate the effective ionization threshold for this lowest wavelength by using Eq. \ref{eq:Flimit}, which gives \SI{6e4}{\meter\per\second}. Integrating the reference $v_x$-velocity distribution found in the previous paragraph for all values that are greater than this threshold and normalizing the result to the total integral, one obtains an estimate for the maximum expected self-ionization signal which resulted in \SI{97}{\percent}.

\begin{figure}[thpb]
	\centering
	\includegraphics[width=1 \linewidth]{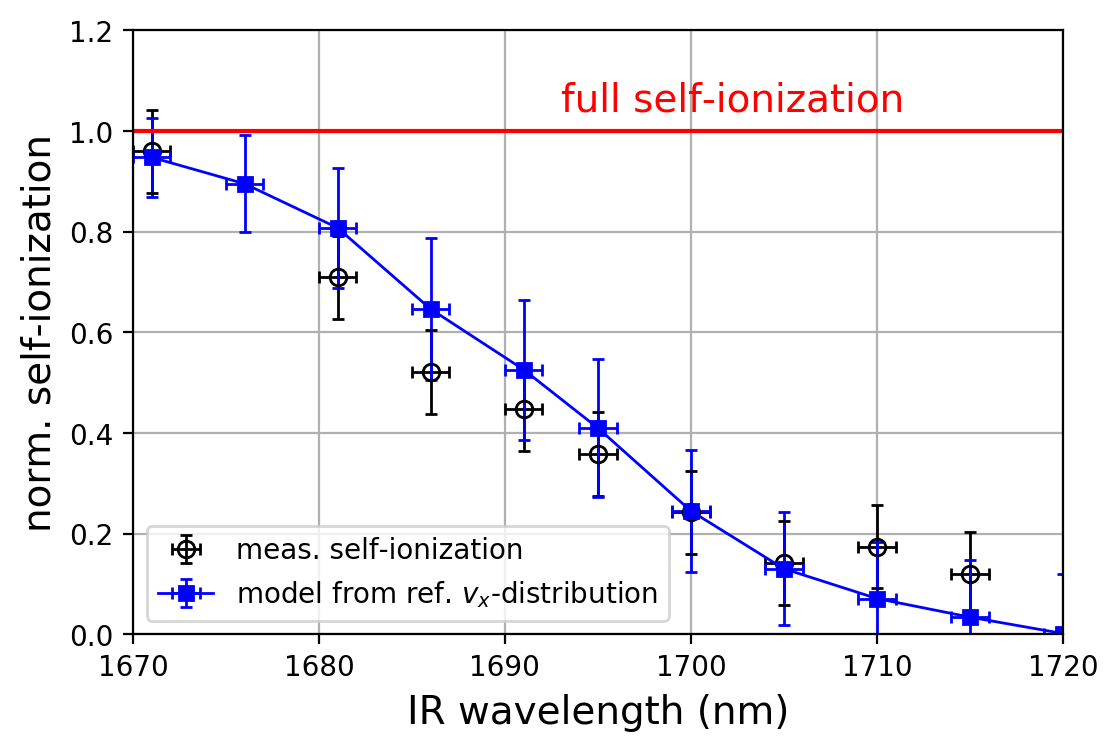}
	\caption{IR-wavelengths scan corresponding to \Ps{}-Rydberg states (circles with error-bars). The self-ionizing fractions have been normalized to the expected full self-ionization. For $\lambda=\SI{1680}{\nano\meter}$, around \SI{70}{\percent} of \Ps{} ionizes. At $\lambda=\SI{1700}{\nano\meter}$, which corresponds to $n_{eff}\approx16$, only \SI{25}{\percent} is lost. 
	The squares are the result of a modeling of the self-ionizing fraction, as detailed in the text. The model resembles the measured self-ionization per effective state.}
	\label{fig:Rydberg-scan}
\end{figure}

The outcome of a self-ionization scan for a fixed delay of \SI{25}{\nano\second} and for the UV-laser set to our reference wavelength is shown in Fig. \ref{fig:Rydberg-scan} (circles with error bars). For $\lambda=\SI{1700}{\nano\meter}$ (i.e. $n_{eff}\approx16$), only about \SI{25}{\percent} ionizes, thus this state seems a reasonable choice when optimizing the charge exchange reaction, as the visible self-ionizing Ps is still very dim, while $n_{eff}$ is not too small. 

The velocimetry result of Fig. \ref{fig:velo_distribution} obtained for the window at \SI{5.05}{\milli\meter} has been used to model the self-ionizing fraction of Rydberg-Ps. For each IR-wavelength, the threshold of self-ionization and the corresponding limiting velocity $v_{x,\:limit}$ have been calculated from Eq. \ref{eq:Flimit}. Then, the fraction of self-ionizing positronium has been computed via numerical integration: all bins with a velocity higher than the ionization-limit contribute to a signal, which is normalized to the result of an integration over all velocities. The model (squares with error-bars) is plotted together with the measured self-ionization over the wavelengths in Fig. \ref{fig:Rydberg-scan}. The model follows roughly the course of the measured self-ionization, which is an indication for the quality of the velocity-distribution along the x-axis.

\subsection{Expected impact on the \Hbar{}-production cross-section}

The cross-section for \Hbar{}-production via the charge-exchange reaction was studied theoretically by different groups. Classical simulations \cite{krasnicky_pra:16} based on Monte-Carlo numerical experiments have given insight into the process for Rydberg states up to $n = 50$. Quantum effects have also been studied in simulations up to $n = 5$, but subsequent investigations have pointed out that the scaling laws are identical \cite{Kadyrov:18,aegis_XSGemma:19}.
Within the framework of the classical Monte-Carlo approach, the predicted cross-section gets larger with increasing Ps principal quantum number $n$ (proportional to $n^2$ at very small velocities and to $n^4$ at intermediate ones), and decreases monotonically with the Ps velocity, becoming negligible at the fast end. 

  \begin{figure}[thpb]
      \centering
      \includegraphics[width=1 \linewidth]{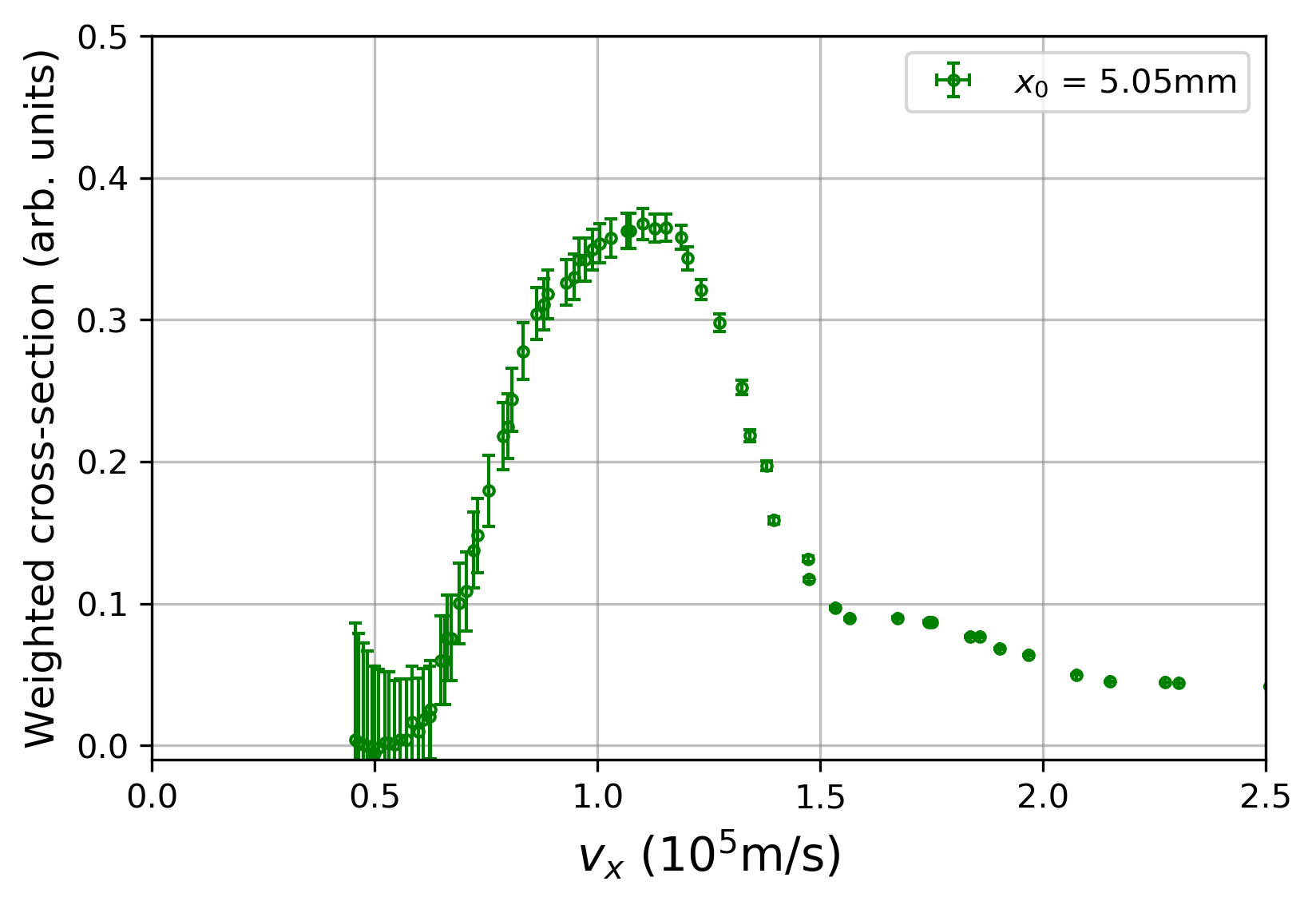}
      \caption{Weighted cross-section for antihydrogen production using Rydberg-\Ps{} with $\lambda_{IR}=\SI{1700}{\nano\meter}$, plotted over the whole accessible range of Ps velocities along the x-axis.}
    \label{fig:Hbar_effi}
    \end{figure}

By using these studies, one can optimize the described Rydberg-Ps source for future efficient antihydrogen-production in \aegis{}.
For this purpose, we weighted the classical cross-section with the measured reference $v_x$-velocity distribution of Ps.
This has been done for all accessible IR-wavelengths. The result for $\lambda_{IR}=\SI{1700}{\nano\meter}$ is shown in \fig{Hbar_effi}. The weighted cross-section shows a plateau in the region of \SI{1e5}{\meter\per\second}, which is due to the strong suppression of faster components, where the relative velocity between Ps and antiprotons is missmatched. 
When properly tuning the laser pulses' delay and thus the selected range of velocities, one can maximize the weighted cross-section (with an optimal velocity $v_{best}$). More so, as it is expected to progressively increase with decreasing $\lambda_{IR}$.

    \begin{figure}[thpb]
    	\centering
    	\includegraphics[width=1 \linewidth]{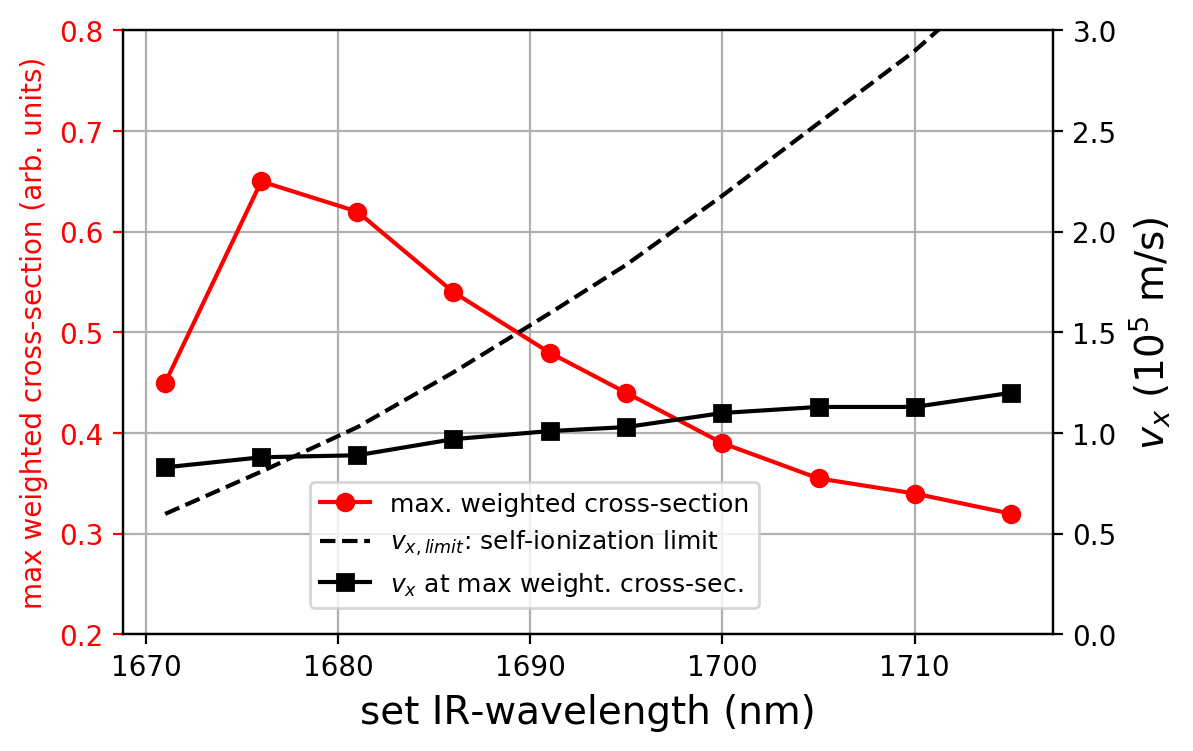}
    	\caption{On the left ordinate: maximum of the weighted cross-sections; on the right ordinate: the vertical velocity-component $v_x$; both plotted over the set IR-laser wavelength addressing the effective Rydberg-state. The solid lines are meant as eye-guides, while the dashed line represents the limiting velocity of self-ionization.}
    	\label{fig:HbaroverN}
    \end{figure}
    
If $v_{best}$ is smaller than the limiting velocity $v_{x,limit}$ for a given IR-wavelength, it is convenient to address positronium with that velocity to have the highest weighted cross-section for \Hbar{}-production. On the other side, if $v_{best}$ is greater than the limiting velocity, positronium travelling with that velocity would self-ionize and cannot be used. In this case, Ps with a velocity regime slightly before the limit should be addressed by an appropriate choice of laser delay.

In Fig. \ref{fig:HbaroverN} the behaviour of $v_{best}$ and of $v_{x,limit}$ is shown as a function of the IR-wavelength. For $\lambda_{IR}\geq\SI{1680}{\nano\meter}$, $v_{best}$ is lower than $v_{x,limit}$.  For example, the weighted cross-section gains roughly a factor of two when going from \SIrange{1700}{1676}{\nano\meter} with our Ps-source, despite the dramatic increase in self-ionization. The improvement originates from the scaling of the surviving fraction with $n^4$. 
Additionally, when optimizing the positron implantation energy or possibly the morphology of the target itself, one could obtain a slower Ps velocity-distribution, which potentially allows to reduce the wavelength even more as the self-ionizing fraction is reduced.

\section{CONCLUSIONS}

In this work we have characterized the velocity emission of positronium from a nanochanneled positron/positronium converter in a cryogenic environment within a magnetic field of \SI{1}{\tesla}. The self-ionization of positronium due to motional Stark effect has also been studied as a function of the wavelength of the IR-laser exciting Ps to Rydberg states, with an effective principal quantum number $n_{eff}$ ranging between 14 and 22. The measurements have been performed by means of MCP-imaging of ionized positronium. In the case of velocimetry, positronium was photoionized after excitation, while for the study of the motional Stark-effect, self-ionized Rydberg-positronium has been imaged. The velocimetry was performed for two axis. The velocity-components aligned parallel to the laser beam have been studied by Doppler-scanning the UV-wavelength populating the $n=3$-manifold, the velocity-components perpendicular to the laser and to the magnetic field were measured by timing scans of the delays for the laser pulses. 
The choice of the positronium velocity and Rydberg state of the atom has been discussed in view of their optimization for \Hbar{}-production via charge exchange reaction. 

\appendix{}
\renewcommand{\theequation}{A-\arabic{equation}}
\setcounter{equation}{0}  
\section{APPENDIX} \label{ModelingApp}  
\subsection{Modeling of the Ps $v_x$-distribution}

\begin{figure}[thpb]
	\centering
	\includegraphics[width=1 \linewidth]{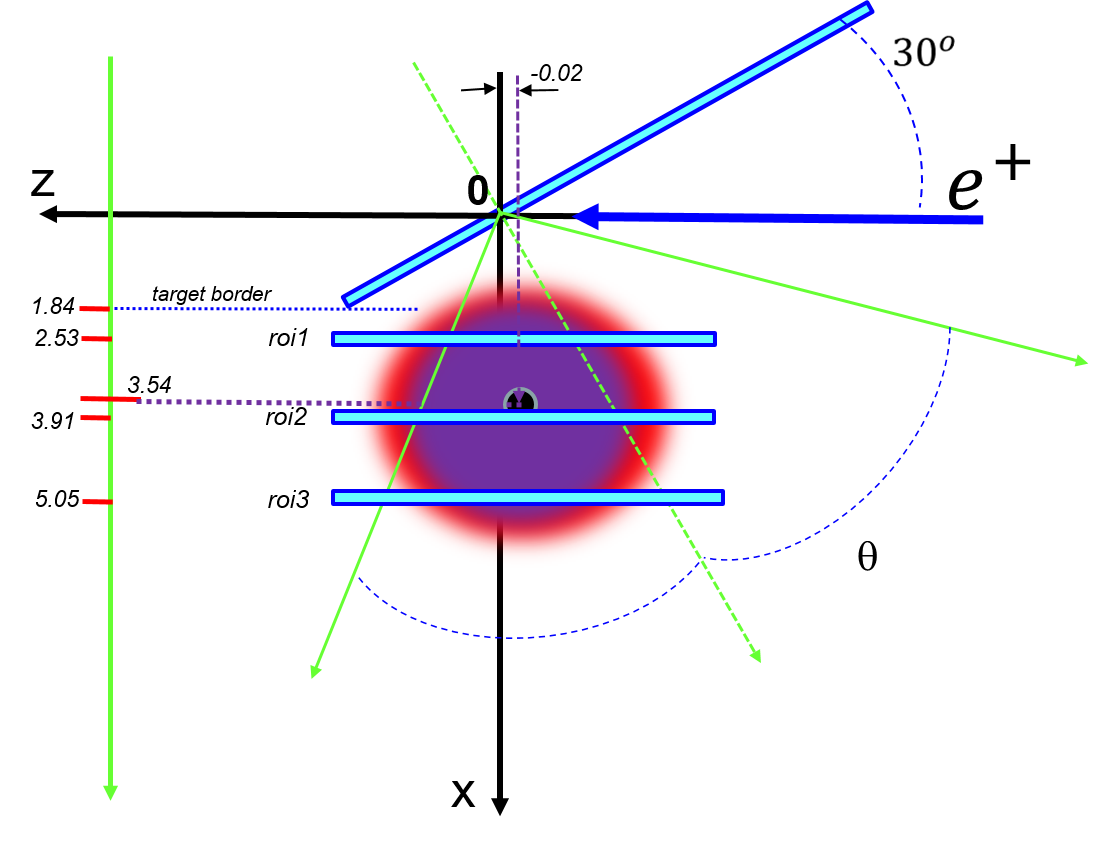}
	\caption{Geometry used for the modeling of Ps photoionization with the three \emph{rois} 1,2,3 as used during the experimental analysis.}
	\label{fig:Model}
\end{figure}

We consider a positron bunch impacting and entering the positron/Ps converter at the reference position (0, 0, 0) of our coordinate system and assign a temporal spread modelled by a normalized Gaussian function centered at time $t=0$ to it:
\begin{equation} \label{A1}
    n_p(t) \, = \frac{1}{\sigma_p \sqrt{2 \pi}}\, e^{-t^2/2 \sigma_p^2}
\end{equation}

For $\sigma_p$ we assume a value of \SI{5}{\nano\second} as was estimated from experimental data. Every positron in the bunch is then assumed to form Ps, which exits the converter from the reference point after an averaged Ps permanence time $t_{perma}$ inside the target. Hence, we get an initial Ps-cloud with the same time-distribution as the positron bunch, but centered at $t=t_{perma}$. We then also define a velocity distribution function $f_0(v_x,v_y,v_z)$ for the Ps-cloud, which we model with a Gaussian function as well, centered at $v_{x,y,z}=\SI{0}{\meter\per\second}$.

The possible resulting velocity-vectors are limited to an emission cone with its vertex located at the Ps-origin and a half opening-angle $\theta$ from the target normal (see Fig. \ref{fig:Model}). The Ps cloud is freely expanding in space obeying the linear law:
\begin{equation}  \label{A2}
x(t_f) = v_x \, t_f \, , \;\;\;\; y(t_f) = v_y \, t_f \, , \;\;\;\; z(t_f) = v_z \, t_f \, ,
\end{equation}
where $t_f$ is the real Ps flight-time. We can therefore describe the Ps cloud by a spatial distribution function $f(x(t_f),y(t_f),z(t_f))$ within the emission cone. Equating the number of Ps contained in a spatial volume-element $dx\,dy\,dz$ with the number of Ps contained in a velocity volume-element $dv_x\, dv_y\, dv_z$ we can express the spatial distribution in terms of the initial velocity distribution $f_0$ for each time $t_f > 0$:
\begin{alignat}{1} \label{A3}
    \begin{aligned} 
    f(x,y,z) \, dx \, dy \, dz \, &= \, f_0(v_x,v_y,v_z) \, dv_x \, dv_y \, dv_z\\
    \Rightarrow f(x,y,z) \, &= \, \frac{1}{t_f^3} \, f_0(v_x,v_y,v_z) \,
    \end{aligned}
\end{alignat}

During their free flight in vacuum, Ps atoms can be probed by two synchronized and spatially overlapped laser pulses (UV+IR, aligned with the y-axis) by means of an initial excitation to $n = 3$ and a subsequent ionization, which we assume for practical reasons to work instantaneously.
The instant of laser-firing corresponds to a time $t = t_f + t_{perma}$. The laser beam profiles are assumed to be Gaussian transverse functions centered in $(x_L, z_L)$, and with $2\Delta r$ as FWHM.
We define a function $\chi_{cone}(\theta)$ which is describing the overlap of the Ps emission cone and the laser beam. If $\theta$ is sufficiently large, which is the case for $\theta\geq\SI{60}{\degree}$, the function $\chi_{cone}(\theta)=1$. 

The number of detected ionized Ps atoms in the \emph{roi} centered at $x_r$ at observation time $t$ is given by a spatial integral:
\begin{widetext}\label{A4}
	\begin{equation}	
    N_r(t)  \equiv  N_r(t_f + t_p)  \, = \,
    g_r \int_{x_r-\Delta x/2}^{x_r+\Delta x/2} dx \,  \int_{-\Delta y/2}^{\Delta y/2} dy \, \int_{z_L-\Delta z}^{z_L+\Delta z} dz \,\, \chi_{cone}(\theta) \times f(x(t_f),y(t_f),z(t_f))
	\end{equation}
\end{widetext}
where  $g_r$ is the efficiency of ionization as was calculated following the procedure described in Ref. \cite{castelli_psexc:08}. The slightly varying values for $g_{1,2,3}$ as given in Tab. \ref{tab:model} are due to the geometrical overlap of the \emph{rois} with the Gaussian spatial laser profile, which at the same time is defining the third integral's boundary, $\Delta z$, via Pythagorean theorem: 
\begin{equation} \label{A5}
    \Delta z = \sqrt{\Delta r^2 - (x_r - x_L)^2} \,\,
    \left\{
        \begin{aligned}
        	\Delta z &\text{  for  } |x_r - x_L| \, \leq \Delta r\\
        	0 &\text{  for  } |x_r - x_L| \, > \Delta r
	    \end{aligned}
	\right.
\end{equation}

From the distribution $N_r(t)$, which is a small excerpt from the initial spatial distribution, we can derive the measurable $v_x-$velocity distribution $P_r(v_x)$ with a simple procedure using the above definitions:

\begin{dmath*} 
	$$	
	P_r(v_x) \, \equiv \, \int \, d v_y  d v_z \,\, f_0\left(v_x, v_y, v_z\right) \, = \, 
	\frac{1}{t_f^2} \, \int \, dy dz \,\, f_0\left(v_x, v_y, v_z\right) \, = \,
	t_f \, \int \, dy dz \,\, f(x(t_f),y(t_f),z(t_f))
	$$
\end{dmath*}

Assuming as before a sufficiently large Ps emission cone, the integrals are limited
by the \emph{rois}' width $\Delta x$ and by the laser beam profile, and with eq. (A-4)
we finally obtain the formula:
\begin{equation}\label{A6}
P_r(v_x) \, \simeq \, \frac{t_f}{g_r \, \Delta x} \, N_r(t) 
\end{equation}

\begin{table}[thpb] 
    \caption{Experimental and assumed parameters for the simulation}
    \label{tab:model}
    \begin{tabular}{|l|c|c|}
         \hline
         $\Delta x $ &  0.46 mm &  \emph{roi}-width along $x$  \\
         $\Delta y $ &  11.5 mm &  \emph{roi}-width along $y$  \\
         $x_1 $ &  2.53 mm &  center of \emph{roi} 1 \\
         $x_2 $ &  3.91 mm &  center of \emph{roi} 2 \\
         $x_3 $ &  5.05 mm &  center of \emph{roi} 3 \\
         $x_L $ &  3.54 mm &  x-position of laser center  \\
         $z_L $ & -0.02 mm & z-position of laser center   \\
         $ \Delta r$ &  2.2 mm & laser spot HWHM \\
         $ g_1$ &  0.55 & efficiency of Ps ionization in \emph{roi} 1 \\
         $ g_2$ &  0.62 & efficiency of Ps ionization in \emph{roi} 2 \\
         $ g_3$ &  0.53 & efficiency of Ps ionization in \emph{roi} 3  \\
         $ \theta$ &  \SI{60}{\degree}& half opening-angle of Ps emission \\
         $ \sigma_{p}$ &  \SI{5}{\nano\second} & temporal spread of $e^+$ bunch \\
         $ \sigma_{v}$ &  $2\times 10^5$ m/s & velocity spread of Ps cloud \\
         $ t_{perma}$ &  \SI{3}{\nano\second} & Ps permanence time \\
         \hline
    \end{tabular}
\end{table}

\begin{figure}[thpb]
	\centering
	\includegraphics[width=1 \linewidth]{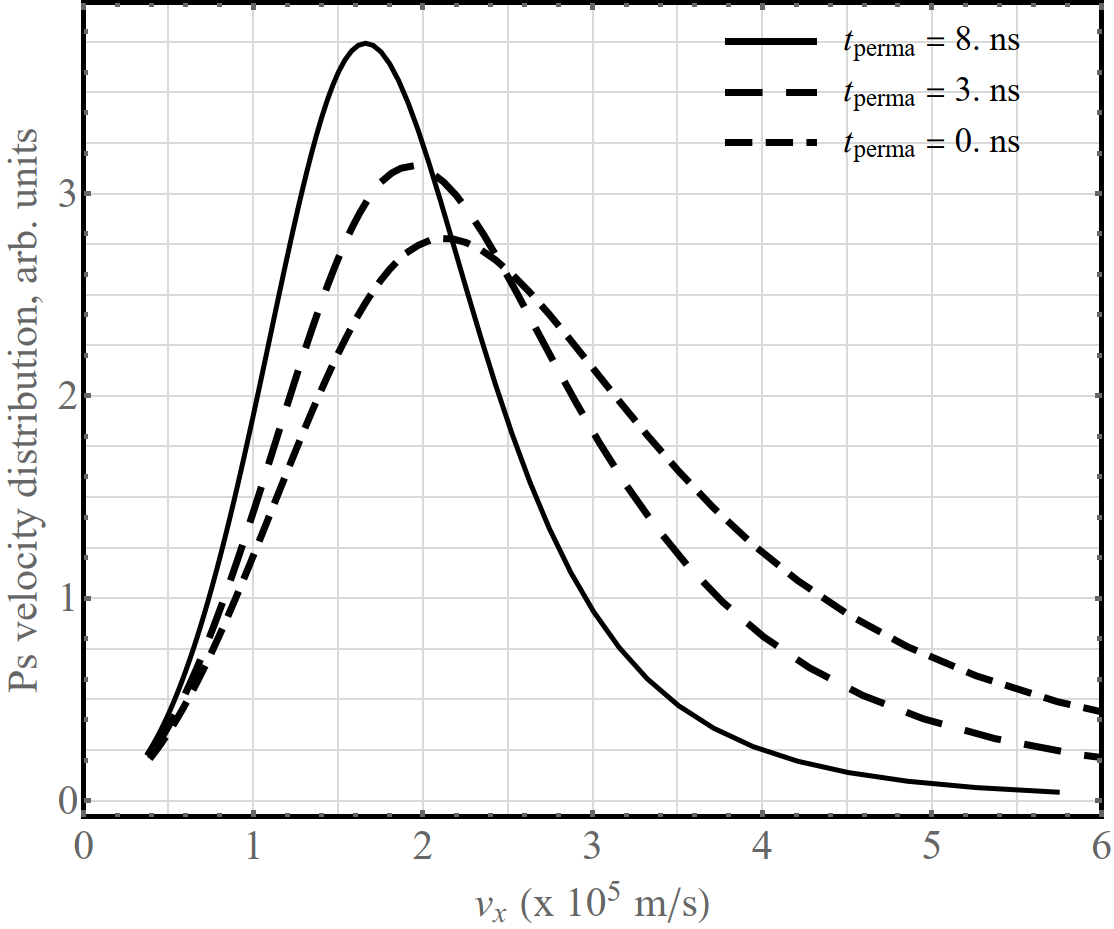}
	\caption{Variance of the modeled $v_x$-velocity distribution for a reasonable range of permanence times $t_{perma}$ in \emph{roi} 3.}
	\label{fig:Model_tp}
\end{figure}

\begin{figure}[thpb]
	\centering
	\includegraphics[width=1 \linewidth]{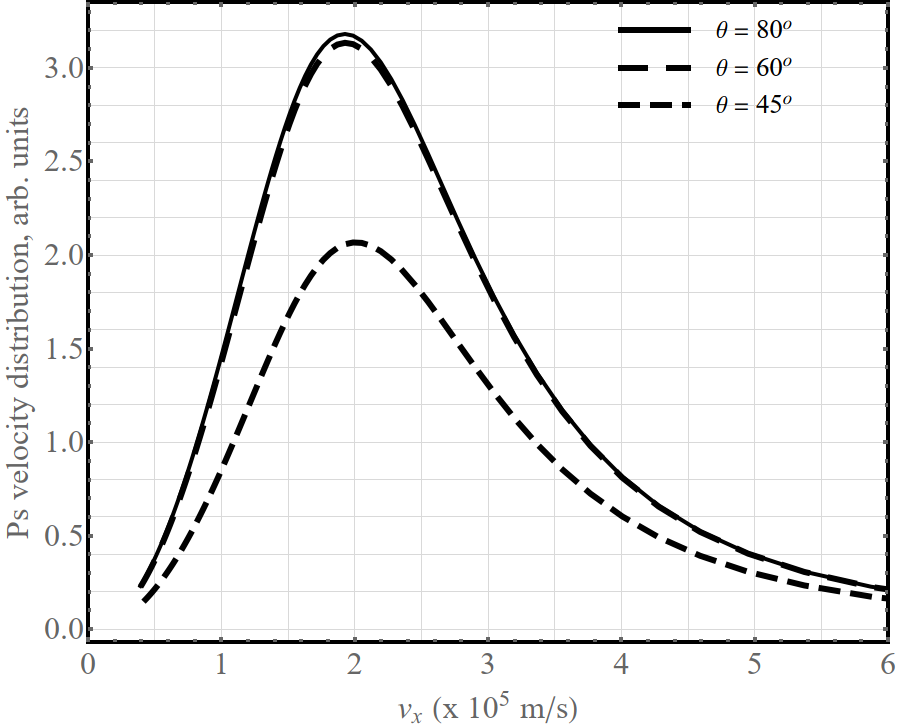}
	\caption{Dependence of the modeled $v_x$-velocity distribution on Ps emission cone defined by $\theta$ in \emph{roi} 3.}
	\label{fig:Model_cone}
\end{figure}

The approximate simulations reported in Fig. \ref{fig:velo_distribution} have been obtained using the parameters in Tab. \ref{tab:model}. We show the influence of the parameters $t_{perma}$ and $\theta$ in Fig. \ref{fig:Model_tp} and Fig. \ref{fig:Model_cone}, where a reasonable range has been explored for our reference \emph{roi} 3. 
The Ps intensity varies less, the peak-velocity more when changing the permanence time $t_{perma}$. As the velocities directly depend on the real time-of-flight, this strong correlation is expected. Varying the Ps emission cone, for angles greater than \SI{60}{\degree} no change is occurring because the laser beam is fully overlapping with the Ps cloud. Reducing that angle, the amplitude decreases rapidly while the peak velocity slightly increases.

\section*{ACKNOWLEDGMENTS}

This work was supported by Istituto Nazionale di Fisica Nucleare; the CERN Fellowship programme and the CERN Doctoral student programme; the Swiss National Science Foundation Ambizione Grant (No. 154833); a Deutsche Forschungsgemeinschaft research grant; an excellence initiative of Heidelberg University; Marie Sklodowska-Curie Innovative Training Network Fellowship of the European Commission's Horizon 2020 programme (No. 721559 AVA); European Research Council under the European Union's Seventh Framework Program FP7/2007-2013 (Grants Nos. 291242 and 277762); European Union's Horizon 2020 research and innovation programme under the Marie Sklodowska-Curie grant agreement ANGRAM No. 748826; the European Union’s Horizon 2020 research and innovation programme under the Marie Sklodowska-Curie Cofund Action, grant agreement No. 754496; Austrian Ministry for Science, Research, and Economy; Research Council of Norway; Bergen Research Foundation; John Templeton Foundation; Ministry of Education and Science of the Russian Federation and Russian Academy of Sciences and the European Social Fund within the framework of realizing research infrastructure for experiments at CERN, LM2015058. 

\bibliographystyle{apsrev}
\bibliography{aegis_bib}

\end{document}

%% file: header.tex
\usepackage{graphicx}
\usepackage[utf8]{inputenc}
\usepackage{siunitx}
\usepackage{amsmath}
\usepackage{amssymb}
\usepackage{enumerate}
\usepackage{color}
\usepackage{hyperref}
\usepackage[american]{babel}
\usepackage{etoolbox}
\usepackage{floatrow}
\usepackage{lineno,hyperref}
\usepackage{subfigure}
\usepackage[disable]{todonotes}
\usepackage{tikz}
\usepackage{nicefrac}
\usepackage{breqn}

\patchcmd{\thebibliography}{\section*{\refname}}{}{}{}

\newcommand{\fig}[1] {Fig.~\ref{fig:#1}}

\newcommand{\aegis}{AE$\bar{\hbox{g}}$IS}
\newcommand{\Hbar}{$\bar{\hbox{H}}$}

\newcommand{\pos}{$\mathrm{e^+}$}

\newcommand{\Ps}{$\rm Ps$}

\newcommand{\macor}[0]{$\text{MACOR}$}



%% file: aegis_authors_list.tex
\newcommand{\corresponding}[1]{\altaffiliation{Corresponding author, #1}}
\newcommand{\correspondingly}[1]{\altaffiliation{\underline{Corresponding author, #1}}}
\newcommand{\affpolimi}[0]{\affiliation{LNESS, Department of Physics, Politecnico di Milano, via Anzani 42, 22100~Como, Italy}}
\newcommand{\affinfnmi}[0]{\affiliation{INFN, Sezione di Milano, via Celoria 16, 20133~Milano, Italy}}
\newcommand{\affvienna}[0]{\affiliation{Stefan Meyer Institute for Subatomic Physics, Austrian Academy of Sciences, Boltzmanngasse 3, 1090~Vienna, Austria}}
\newcommand{\affinsubria}[0]{\affiliation{Department of Science and High Technology, University of Insubria, Via Valleggio 11, 22100~Como, Italy}}
\newcommand{\affjinr}[0]{\affiliation{Joint Institute for Nuclear Research, Dubna~141980, Russia}}
\newcommand{\affbs}[0]{\affiliation{Department of Mechanical and Industrial Engineering, University of Brescia, via Branze 38, 25123~Brescia, Italy}}
\newcommand{\affinfnpv}[0]{\affiliation{INFN Pavia, via Bassi 6, 27100~Pavia, Italy}}
\newcommand{\afftn}[0]{\affiliation{Department of Physics, University of Trento, via Sommarive 14, 38123~Povo, Trento, Italy}}
\newcommand{\affinfntn}[0]{\affiliation{TIFPA/INFN Trento, via Sommarive 14, 38123~Povo, Trento, Italy}}
\newcommand{\affge}[0]{\affiliation{Department of Physics, University of Genova, via Dodecaneso 33, 16146~Genova, Italy}}
\newcommand{\affinfnge}[0]{\affiliation{INFN Genova, via Dodecaneso 33, 16146~Genova, Italy}}
\newcommand{\affmi}[0]{\affiliation{Department of Physics ``Aldo Pontremoli'', Universit\`{a} degli Studi di Milano, via Celoria 16, 20133~Milano, Italy}}
\newcommand{\affmpi}[0]{\affiliation{Max Planck Institute for Nuclear Physics, Saupfercheckweg 1, 69117~Heidelberg, Germany}}
\newcommand{\afflac}[0]{\affiliation{Laboratoire Aim\'e Cotton, Universit\'e Paris-Sud, ENS Paris Saclay, CNRS, Universit\'e Paris-Saclay, 91405~Orsay Cedex, France}}
\newcommand{\affpolimiII}[0]{\affiliation{Department of Aerospace Science and Technology, Politecnico di Milano, via La Masa 34, 20156~Milano, Italy}}
\newcommand{\affheidelberg}[0]{\affiliation{Kirchhoff-Institute for Physics, Heidelberg University, Im Neuenheimer Feld 227, 69120~Heidelberg, Germany}}
\newcommand{\affcern}[0]{\affiliation{Physics Department, CERN, 1211~Geneva~23, Switzerland}}
\newcommand{\affoslo}[0]{\affiliation{Department of Physics, University of Oslo, Sem Saelandsvei 24, 0371~Oslo, Norway}}
\newcommand{\afflyon}[0]{\affiliation{Institute of Nuclear Physics, CNRS/IN2p3, University of Lyon 1, 69622~Villeurbanne, France}}
\newcommand{\affmoscow}[0]{\affiliation{Institute for Nuclear Research of the Russian Academy of Science, Moscow~117312, Russia}}
\newcommand{\affinfnpd}[0]{\affiliation{INFN Padova, via Marzolo 8, 35131~Padova, Italy}}
\newcommand{\affprague}[0]{\affiliation{Czech Technical University, Prague, Brehov\'a 7, 11519~Prague~1, Czech Republic}}
\newcommand{\affbo}[0]{\affiliation{University of Bologna, Viale Berti Pichat 6/2, 40126~Bologna, Italy}}
\newcommand{\affpv}[0]{\affiliation{Department of Physics, University of Pavia, via Bassi 6, 27100~Pavia, Italy}}
\newcommand{\affnorway}[0]{\affiliation{The Research Council of Norway, P.O. Box 564, 1327~Lysaker, Norway}}
\newcommand{\affheidelbergII}[0]{\affiliation{Department of Physics, Heidelberg University, Im Neuenheimer Feld 226, 69120~Heidelberg, Germany}}
\newcommand{\affbsII}[0]{\affiliation{Department of Civil, Environmental, Architectural Engineering and Mathematics, University of Brescia, via Branze 43, 25123~Brescia, Italy}}



\author{M.~Antonello}
\affinfnmi
\affinsubria

\author{A.~Belov}
\affmoscow

\author{G.~Bonomi}
\affbs
\affinfnpv

\author{R.~S.~Brusa}
\afftn
\affinfntn

\author{M.~Caccia}
\affinfnmi
\affinsubria

\author{A.~Camper}
\corresponding{antoine.camper@cern.ch}
\affcern

\author{R.~Caravita}
\afftn

\author{F.~Castelli}
\affinfnmi
\affmi


\author{D.~Comparat}
\afflac


\author{G.~Consolati}
\affpolimiII
\affinfnmi


\author{L.~Di~Noto}
\affge
\affinfnge

\author{M.~Doser}
\affcern

\author{M.~Fan\`{i}}
\affge
\affinfnge
\affcern

\author{R.~Ferragut}
\affpolimi
\affinfnmi

\author{J.~Fesel}
\affcern


\author{S.~Gerber}
\affcern


\author{A.~Gligorova}
\affvienna

\author{L.~T.~Gl\"oggler}
\affcern

\author{F.~Guatieri}
\afftn
\affinfntn


\author{S.~Haider}
\affcern

\author{A.~Hinterberger}
\affcern



\author{O.~Khalidova}
\affcern

\author{D.~Krasnick\'y}
\affinfnge

\author{V.~Lagomarsino}
\affge
\affinfnge



\author{C.~Malbrunot}
\affcern

\author{S.~Mariazzi}
\afftn
\affinfntn

\author{V.~Matveev}
\affmoscow
\affjinr

\author{S.~R.~M\"{u}ller}
\affheidelberg

\author{G.~Nebbia}
\affinfnpd

\author{P.~Nedelec}
\afflyon

\author{L.~Nowak}
\affcern

\author{M.~Oberthaler}
\affheidelberg

\author{E.~Oswald}
\affcern

\author{D.~Pagano}
\affbs
\affinfnpv


\author{L.~Penasa}
\afftn
\affinfntn

\author{V.~Petracek}
\affprague

\author{F.~Prelz}
\affinfnmi


\author{B.~Rien\"acker}
\correspondingly{benjamin.rienaecker@cern.ch}
\affcern


\author{O.~M.~R{\o}hne}
\affoslo

\author{A.~Rotondi}
\affinfnpv
\affpv

\author{H.~Sandaker}
\affoslo

\author{R.~Santoro}
\affinfnmi
\affinsubria



\author{G.~Testera}
\affinfnge

\author{I.~C.~Tietje}
\affcern


\author{T.~Wolz}
\affcern


\author{C.~Zimmer}
\affcern
\affoslo
\affheidelbergII

\author{N.~Zurlo}
\affinfnpv
\affbsII

\collaboration{The AEgIS collaboration}
\noaffiliation{}